\documentclass[twocolumn,pre,floats,aps,amsmath,amssymb]{revtex4}
\usepackage{graphicx}
\usepackage{bm}
\usepackage{mathtools}
\usepackage{mathrsfs}
\usepackage{hyperref}
\usepackage[dvipsnames]{xcolor}
\usepackage{amsmath}
\usepackage{caption}
\usepackage{physics}

\begin{document}

\title{Particle Production by a Relativistic Semi-transparent Mirror 
\\
in 1+3D Minkowski Spacetime}
\author{Kuan-Nan Lin,${}^{1,2}$\footnote{r08222037@ntu.edu.tw} Chih-En Chou,${}^{1,2}$\footnote{r07222028@ntu.edu.tw} and Pisin Chen${}^{1,2,3}$\footnote{pisinchen@phys.ntu.edu.tw}}
\affiliation{
${}^{1}$Leung Center for Cosmology and Particle Astrophysics,
National Taiwan University, Taipei 10617, Taiwan, ROC
\\
${}^{2}$Department of Physics and Center for Theoretical Sciences, National Taiwan University,
Taipei 10617, Taiwan, ROC
\\
${}^{3}$Kavli Institute for Particle Astrophysics and Cosmology,
SLAC National Accelerator Laboratory,
Stanford University, Stanford, California 94305, USA}
\date{\today}

\begin{abstract}
Production of scalar particles by a relativistic, semi-transparent mirror in 1+3D Minkowski spacetime based on the Barton-Calogeracos (BC) action is investigated. The corresponding Bogoliubov coefficients are derived for a mirror with arbitrary trajectory. In particular, we apply our derived formula to the gravitational collapse trajectory. In addition, we identify the relation between the particle spectrum and the particle production probability, and we demonstrate the equivalence between our approach and the existing approach in the literature, which is restricted to 1+1D. In short, our treatment extends the study to 1+3D spacetime. Lastly, we offer a third approach for finding the particle spectrum using the S-matrix formalism.
\end{abstract}

\maketitle

\section{Introduction}
\label{sec:intro}

In 1970, Moore demonstrated \cite{0} that quanta of electromagnetic field may be produced from the initial vacuum state if the field is constrained in a one-dimensional cavity and subject to time-dependent Dirichlet boundary conditions in 1+1D Minkowski spacetime. This phenomenon is a manifestation of the interaction between vacuum fluctuations of the quantized field and moving boundaries. A few years later, DeWitt \cite{0a} showed that, for a scalar field subject to a single time-dependent Dirichlet boundary condition, i.e., a perfect moving point mirror, in 1+1D Minkowski spacetime, the production of particles out of the initial vacuum state is also possible. Soon after, Fulling and Davies studied the energy-momentum tensor \cite{4} and particle spectrum \cite{4a} for a perfect point mirror following prescribed trajectories in 1+1D. The production of particles out of the vacuum due to time-dependent boundary condition(s) is therefore referred to as the: "Moore effect", "Dynamical Casimir effect", "Motion-induced radiation", or "Moving mirror model". For mirrors with a variety of trajectories mimicking different scenarios of black hole radiation, please see Good's recent works, e.g., \cite{4g}\cite{4h}, whereas for various trajectories mimicking different candidate resolutions to the information loss paradox of black hole evaporation, please see Chen and Yeom \cite{4i}.

Most work in this subject are studied in 1+1D Minkowski spacetime where the scalar field and the Klein-Gordon equation satisfy the conformal invariance. Conformal invariance allows for exact solutions to the Klein-Gordon equation for a perfect point mirror in arbitrary motion, i.e., arbitrary time-dependent Dirichlet boundary condition. In addition, when expanding the scalar field in terms of mode functions, the past null infinity $\mathscr{I}^{-}$ and the future null infinity $\mathscr{I}^{+}$ can always serve as the in-region and the out-region, respectively, and thus the concept of particle may be defined in these two regions. Nevertheless, conformal invariance breaks down in higher dimensions and thus the techniques developed for 1+1D no longer apply. Instead, in 1+3D spacetime, the proper in-region and the out-region are, respectively, the remote past ($x_0\rightarrow-\infty$) and the remote future ($x_0\rightarrow\infty$). In this case, particle spectra for a non-relativistic mirror with bounded motions starting and ending at the same position have been worked out \cite{2c}\cite{2d}\cite{2g}\cite{2h}\cite{2i} based on the perturbative approach of Ford and Vilenkin \cite{4f}.

Aside from the concept of particles, another physical quantity of common interest is the (local) energy-momentum tensor. This quantity may be much easier to obtain than the particle spectrum (if it may be defined) for mirrors with arbitrary trajectories and even in higher dimensional spacetimes since it only requires the knowledge of the in-mode and the in-vacuum. Energy-momentum tensor is, in general, not related to the particle spectrum by a simple mode-summation procedure of adding up the energy carried by each particle, see, e.g., \cite{4}\cite{4a}\cite{4g}\cite{4h}\cite{4f}\cite{4e}, because it also contains the effect of vacuum polarization. Energy-momentum tensor for an infinite-size, plane, Rindler mirror in 1+3D has been worked out by Candelas and Raine \cite{4c} and Candelas and Deutsch \cite{4d} while spherical mirrors with nearly-uniform acceleration are also studied in, e.g., \cite{5}\cite{5a}\cite{5b}.

Limited by technologies, a direct construction of a relativistic mirror in laboratories to test the above studies were not feasible. Therefore, alternative experimental proposals had been conceived and conducted, e.g., the superconducting quantum interference device (SQUID) experiment \cite{7} and references therein. Nevertheless, it is recently proposed by Chen and Mourou \cite{7a}\cite{7b} that the relativistic mirror may be manifested through plasma wakefields.

In actual experiments, such as that proposed in \cite{7a}\cite{7b} that involve physical mirrors, one tends to encounter the following situations: (i) the mirror is not a perfect reflector, (ii) the spacetime is 1+3D, and (iii) the mirror is finite in size. Therefore, formulations that incorporate these realistic, less than perfect situations are desirable for the cross-check with experimental results. In this paper, the approach we adopted, in principle, enables us to include these situations. However, in this paper, we will only focus on (i) and (ii).

We will begin with the Barton-Calogeracos (BC) action \cite{6a}:
\begin{equation}
\begin{aligned}
S_{\alpha}[\phi]&=-\frac{1}{2}\int_{\mathbb{R}} d^4x\;\partial^{\mu}\phi(x)\partial_{\mu}\phi(x)
\\
&\quad-\frac{\alpha}{2}\int_{\mathbb{R}} d^4x\gamma^{-1}(x_0)\delta(x_3-q(x_0))\phi^2(x)\;,
\end{aligned}
\end{equation}
where $\alpha$ is a coupling constant with the dimension of length${}^{-1}$, $\gamma(x_0)$ is the usual Lorentz factor, and $q(x_0)$ denotes the mirror's trajectory. In this action, the scalar field $\phi(x)$ interacts quadratically with a Dirac-delta function that simulates the moving mirror. Their interaction is adiabatically switched on and off in the remote past $(x_0\rightarrow-\infty)$ and the remote future $(x_0\rightarrow\infty)$ and thus we will identify them as the in-region and the out-region, respectively. The BC action is applicable to relativistic, partial reflecting mirrors and general spacetime dimensions. The model is equivalent to a jellium sheet of zero width, i.e., a surface of zero thickness with a surface current density generated by the motion of small charge elements with charge density $n_s$ (number of charge elements per unit area) and the coupling constant is identified as $\alpha=2\pi n_s e^2/m_e$, where $e$ and $m_e$ are the charge and the mass of the individual entity, respectively \cite{6a}\cite{6}.

Despite the generalizability of the BC action, so far only reductions to 1+1D or the non-relativistic limit have been studied, e.g., \cite{1}-\cite{1h}. Recently, Fosco, Giraldo and Mazzitelli \cite{2f} studied the pair production probability for the BC action in higher dimensional spacetime by using the in-out effective action approach. In this paper, we (i) derive the particle spectrum for a mirror following general, prescribed trajectories by solving the inhomogeneous Klein-Gordon equation for the BC action using the Born approximation and subsequently (ii) identify the relation between the particle spectrum and the particle production probability. In addition, we demonstrate the equivalence of our approach to Nicolaevici's approach \cite{1d}\cite{1e} in 1+1D.

This paper is organized as follows. In Sec.~\ref{sec:II}, the 1+3D inhomogeneous Klein-Gordon equation is solved by using the Born approximation. The Bogoliubov transformation between the in- and the out-annihilation and creation operators are subsequently derived and the particle spectrum follows straightforwardly. In addition, the relation between the particle spectrum and the particle production probability is found. In Sec.~\ref{sec:III}, we demonstrate the equivalence between our treatment and the approach adopted in the literature in 1+1D. In Sec.~\ref{sec:IV}, we apply our 1+3D formula to the gravitational collapse trajectory. The conclusion is given in Sec.~\ref{sec:V}. In Appendix~\ref{A1}, we offer a third approach for finding the particle spectrum using the S-matrix formalism.

In this paper, we use units in which $\hbar=k_B=c=1$ and the signature of the metric tensor is $(-,+,+,+)$ in 1+3D. The symbol $x$ refers to $(x_0,\textbf{x}_{\perp},x_3)$, where $\textbf{x}_{\perp}$ are the coordinates $x_1,x_2$ that are transverse to the mirror's motion. Last but not least, $\mathbb{R}$ refers to $(-\infty,\infty)$ and $\mathbb{R}^{+}$ refers to $(0,\infty)$, etc.

\section{Particle Production in 1+3D}
\label{sec:II}
\subsection{Particle Spectrum}

The equation of motion (EOM) for the BC action is
\begin{equation}
\partial^{\mu}\partial_{\mu}\phi(x)=\alpha\gamma^{-1}(x_0)\delta(x_3-q(x_0))\phi(x)\;.
\end{equation}
Due to the linearity of the differential equation, its \textit{solution}\footnote{This is an integral equation that can be solved iteratively.} can be superposed by
\begin{equation}
\phi(x)=\phi_{h}(x)+\phi_{p}(x)\;,
\end{equation}
where (after second-quantization)
\begin{equation}
\begin{aligned}
&\hat{\phi}_h(x)
=\int\frac{d^3k'}{(2\pi)^{3/2}(2|\textbf{k}'|)^{1/2}}\left[\hat{a}_{\textbf{k}'}e^{-i|\textbf{k}'|x_0+i\textbf{k}'\cdot\textbf{x}}+\text{H.c} \right],
\end{aligned}
\label{III3}
\end{equation}
is the homogeneous solution with its integration range to be determined and
\begin{equation}
\hat{\phi}_p(x)=-\alpha\int_{\mathbb{R}} d^4x'\gamma^{-1}(x_0')\delta(x_3'-q(x_0'))\hat{\phi}(x')G_{R}(x,x')\;,
\label{III4}
\end{equation}
is the particular \textit{solution}; $G_{R}(x,x')$ is the free field retarded Green function. Applying the Born approximation to the first order in $\alpha$, we obtain
\begin{equation}
\begin{aligned}
&\hat{\phi}^{(1)}(x)=\hat{\phi}_{h}(x)+\hat{\phi}_{p}^{(1)}(x)
\\
&=\hat{\phi}_{h}(x)
\\
&
-\alpha\int_{\mathbb{R}} d^4x'\gamma^{-1}(x_0')\delta(x_3'-q(x_0'))\hat{\phi}_h(x')G_{R}(x,x')\;,
\end{aligned}
\label{III5}
\end{equation}
where the homogeneous solution is now
\begin{equation}
\begin{aligned}
&\hat{\phi}_h
=\int_{\mathbb{D}}\frac{d^3k'}{(2\pi)^{3/2}(2|\textbf{k}'|)^{1/2}}\left[\hat{a}_{\textbf{k}'}e^{-i|\textbf{k}'|x_0+i\textbf{k}'\cdot\textbf{x}}+\text{H.c} \right].
\end{aligned}
\end{equation}

The domain for the integration over momentum is finally determined as $\textbf{k}'\in\mathbb{D}$ by the semi-transparent condition: $|\phi_{p}^{(1)}(x)|\ll |\phi_{h}(x)|$ due to the first-order approximation made. This constraint would lead to a lower momentum cutoff for the incoming free modes. Physically speaking, the modes within $\mathbb{D}$ are those having enough momenta such that the mirror acts semi-transparently.

The counterpart of $\phi^{(1)}(x)$ using the free field advanced Green function, $G_{A}(x,x')$, is similarly obtained:
\begin{equation}
\begin{aligned}
&\hat{\phi}^{(1)}(x)=\hat{\phi}_{h}(x)
\\
&-\alpha\int_{\mathbb{R}} d^4x'\gamma^{-1}(x_0')\delta(x_3'-q(x_0'))\hat{\phi}_h(x')G_{A}(x,x')\;.
\end{aligned}
\label{III6}
\end{equation}

Using the Green functions, we are in fact assuming the vacuum state to be defined by $\hat{a}_{\textbf{k}}^{\text{in}}\left|0,\text{in}\right>=0$. In addition, since we are considering first order solutions, we may identify the homogeneous part of \eqref{III6} as the out-field while the remaining $\hat{\phi}_h$ in \eqref{III5} and \eqref{III6} as the in-field.

To obtain the knowledge of annihilation and creation operators, we Fourier transform $\hat{\phi}^{(1)}(x)$ by
\begin{equation}
\int_{\mathbb{R}} dx_0d^2x_{\perp}\int_{0}^{\infty}dx_3\;\hat{\phi}^{(1)}(x)e^{i\omega x_0-i\textbf{k}\cdot\textbf{x}}\;,
\label{III7}
\end{equation}
and use the Green functions of the following form
\begin{equation}
\begin{aligned}
&G_{R/A}(x,x')=\int_{\mathbb{R}}\frac{d\omega}{2\pi}\frac{e^{-i\omega(x_0-x_0')\pm i\omega|\textbf{x}-\textbf{x}'|}}{4\pi|\textbf{x}-\textbf{x}'|}\;,
\end{aligned}
\label{III8}
\end{equation}
where the Weyl identity for $\omega\in\mathbb{R}^{+}$
\begin{equation}
\begin{aligned}
&\frac{e^{i\omega|\textbf{x}-\textbf{x}'|}}{4\pi|\textbf{x}-\textbf{x}'|}
=\frac{i}{8\pi^2}\int_{\mathbb{R}} d^2k_{\perp}\frac{e^{i\textbf{k}_{\perp}\cdot(\textbf{x}_{\perp}-\textbf{x}_{\perp}')+i(\omega^2-k_{\perp}^2)^{1/2}|x_3-x_3'|}}{(\omega^2-k_{\perp}^2)^{1/2}},
\end{aligned}
\end{equation}
is to be used in the calculation. To proceed with computational ease, we temporary assume $q(x_0')\leq 0\;\forall\; x_0'$ since $x_3$ is already positive in \eqref{III7}. Finally, by equating the Fourier transform of \eqref{III5} and \eqref{III6} and subsequently choosing $\omega>0,\;(\omega^2-k_{\perp}^2)^{1/2}>0,\;\text{and } k_3=(\omega^2-k_{\perp}^2)^{1/2}>0$ after lengthy calculations, we obtain the Bogoliubov transformation on the mirror's right as
\begin{equation}
\begin{aligned}
&\hat{a}_{\textbf{k}_{\perp}k_3}^{\text{out}}\approx\hat{a}_{\textbf{k}_{\perp}k_3}^{\text{in}}+\frac{\alpha}{4\pi i}\frac{1}{|\textbf{k}|^{1/2}}\int_{\mathbb{R}} dx_0'\int_{\mathbb{D}}dk_3'\frac{\gamma^{-1}(x_0')}{(k_{\perp}^2+{k_3'}^2)^{1/4}}
\\
&\times
\biggl[\hat{a}_{\textbf{k}_{\perp}k_3'}^{\text{in}}\;e^{-i(\sqrt{k_{\perp}^2+{k_3'}^2}-|\textbf{k}|)x_0'+i(k_3'-k_3)q(x_0')}
\\
&\quad+\hat{a}_{-\textbf{k}_{\perp}k_3'}^{\text{in}\dagger}\;e^{i(\sqrt{k_{\perp}^2+{k_3'}^2}+|\textbf{k}|)x_0'-i(k_3'+k_3)q(x_0')} \biggr]\;,
\end{aligned}
\end{equation}
where
\begin{equation}
\begin{aligned}
\beta_{\textbf{k}\textbf{k}'}&\approx\frac{\alpha}{4\pi i}\frac{1}{|\textbf{k}|^{1/2}}\int_{\mathbb{R}} dx_0'\frac{\gamma^{-1}(x_0')}{(k_{\perp}^2+{k_3'}^2)^{1/4}}
\\
&\quad\times e^{i(\sqrt{k_{\perp}^2+{k_3'}^2}+|\textbf{k}|)x_0'-i(k_3'+k_3)q(x_0')}\;,
\label{III11}
\end{aligned}
\end{equation}
is our desired beta-coefficient. Similarly, choosing $\omega>0,\;(\omega^2-k_{\perp}^2)^{1/2}>0\;\text{and } k_3=-(\omega^2-k_{\perp}^2)^{1/2}<0$, one obtains the same expression of Bogoliubov transformation as above but with $k_3$ now being negative (this corresponds to the mirror's left).

The number of particles with $\textbf{k}\in\mathbb{D}$ per mode in the out-region is thus
\begin{equation}
\begin{aligned}
&\frac{dN}{d^2k_{\perp}dk_3}=\left<0,\text{in}|\hat{a}_{\textbf{k}_{\perp}k_3}^{\text{out}\dagger}\hat{a}_{\textbf{k}_{\perp}k_3}^{\text{out}}|0,\text{in}\right>\\
&=\frac{A}{4\pi^2}\int_{\mathbb{D}} dk_3'\;|\beta_{\textbf{k}\textbf{k}'}|^2
\\
&\approx\frac{A\alpha^2}{64\pi^4|\textbf{k}|}\int_{\mathbb{D}} dk_3'\frac{1}{\sqrt{k_{\perp}^2+{k_3'}^2}}
\\
&\times
\left|\int_{\mathbb{R}} dx_0'\gamma^{-1}(x_0')e^{i(|\textbf{k}|+\sqrt{k_{\perp}^2+{k_3'}^2})x_0'-i(k_3+k_3')q(x_0')}\right|^2,
\label{III12}
\end{aligned}
\end{equation}
where $A$ is the area of the infinite-size plane mirror. The beta-coefficient and particle spectrum in 1+1D follow directly from \eqref{III11} and \eqref{III12} by letting $\textbf{k}_{\perp}=0$.

\subsection{Particle Production Probability}

The vacuum persistence amplitude $\mathcal{Z}_{\alpha}$ corresponding to the BC action is defined as
\begin{equation}
\mathcal{Z}_{\alpha}=e^{iW_{\alpha}}=\int\mathcal{D}\phi\;e^{iS_{\alpha}[\phi]}\;,
\end{equation}
where $W_{\alpha}$ is the effective action. By decomposing $W_{\alpha}$ as $W_{\alpha}=W_{0}+W_{I}$, where $W_0$ is the effective action in the absence of interaction, i.e., free field effective action, the interaction effective action $W_{I}$ can be written as
\begin{equation}
\begin{aligned}
e^{iW_{I}}&=\left<0\right|\mathcal{T}e^{-\frac{i\alpha}{2}\int_{\mathbb{R}} d^4x\gamma^{-1}(x_0)\delta(x_3-q(x_0))\hat{\phi}^2(x)}\left|0\right>,
\end{aligned}
\end{equation}
where $\mathcal{T}$ is the time-ordering operator, $\hat{\phi}$ is a free scalar field operator, and $\left|0\right>$ is the free field vacuum state. By expanding to the second order in $\alpha$ and using Wick's theorem, we obtain
\begin{equation}
\begin{aligned}
&e^{iW_{I}}\approx 1+\frac{\alpha}{2}\int_{\mathbb{R}} d^4x\gamma^{-1}(x_0)\delta(x_3-q(x_0))G_F(x,x)
\\
&
+\frac{\alpha^2}{8}\left[\int_{\mathbb{R}} d^4x \gamma^{-1}(x_0)\delta(x_3-q(x_0))G_F(x,x)\right]^2
\\
&
+\frac{\alpha^2}{4}\int_{\mathbb{R}} d^4x d^4x'\gamma^{-1}(x_0)\delta(x_3-q(x_0))
\\
&\times\gamma^{-1}(x_0')\delta(x_3'-q(x_0'))G_{F}^2(x,x')\;,
\end{aligned}
\end{equation}
where $G_{F}(x,x')$ is the free field Feynman propagator. The constant factors in the denominator of each term are the symmetry factors for the corresponding processes. For example, the symmetry factor 2 for the $\mathcal{O}(\alpha)$ process comes from the propagator starting and ending on the same spacetime point (vertex); the factor $4=2\times 2^{1}\times 1!$ for the last term originates, respectively, from (i) two propagators connecting $x$ and $x'$, (ii) $2^{2/2}=2^{1}$ ways of choosing $2/2=1$ vertex among the 2 vertices as an in vertex, and (iii) $1!$ way to pair the in vertex with the remaining (out) vertex. $W_{I}$ is approximately
\begin{equation}
\begin{aligned}
&iW_{I}\approx\frac{\alpha}{2}\;G_F(0)A\int_{\mathbb{R}} d\tau
+\frac{\alpha^2}{4}\int_{\mathbb{R}} d^4x d^4x'\gamma^{-1}(x_0)
\\
&\times
\delta(x_3-q(x_0))\gamma^{-1}(x_0')\delta(x_3'-q(x_0'))G_{F}^2(x,x')\;,
\end{aligned}
\end{equation}
where we have used $\ln(1+x)\approx x-x^2/2$ and $\tau$ is the mirror's proper time. Using the following expression for the Feynman propagator
\begin{equation}
\begin{aligned}
&G_{F}(x,x')=-i\Theta(\Delta x_0)  \int_{\mathbb{R}}\frac{d^3k}{(2\pi)^3}\frac{e^{-i\vert \mathbf{k}\vert\Delta x_0+i\mathbf{k}\cdot\Delta\mathbf{x}}}{2\vert \mathbf{k}\vert}
\\
&-i\Theta(-\Delta x_0) \int_{\mathbb{R}}\frac{d^3k}{(2\pi)^3}\frac{e^{i\vert \mathbf{k}\vert\Delta x_0+i\mathbf{k}\cdot\Delta\mathbf{x}}}{2\vert \mathbf{k}\vert}\;,
\end{aligned}
\end{equation}
where $\Delta x_0=x_0-x_0',\; \Delta\textbf{x}=\textbf{x}-\textbf{x}'$,
and replacing $\Theta(-\Delta x_0)$ by $1-\Theta(\Delta x_0)$, we obtain the probability of particle production as
\begin{equation}
\begin{aligned}
&\mathcal{P}\approx 2\text{Im} W\approx \frac{1}{2}\int_{\mathbb{D}}d^3k\frac{A\alpha^2}{64\pi^4|\textbf{k}|}\int_{\mathbb{D}} dk_3'\frac{1}{\sqrt{k_{\perp}^2+{k_3'}^2}}
\\
&\times\left|\int_{\mathbb{R}} dx_0'\gamma^{-1}(x_0')e^{i(|\textbf{k}|+\sqrt{k_{\perp}^2+{k_3'}^2})x_0'-i(k_3+k_3')q(x_0')}\right|^2,
\end{aligned}
\end{equation}
where the factor of $1/2$ in the front of the rhs is the product of $2\times 1/4$. Note that the domains for the momenta are $\mathbb{D}$ since we are only considering the case of a semi-transparent mirror, i.e., second order in $\alpha$. Finally, by comparing with \eqref{III12}, we see that the probability of particle production is related to the particle spectrum by
\begin{equation}
\mathcal{P}\approx 2\text{Im} W\approx\frac{1}{2}\int_{\mathbb{D}}d^3k\frac{dN}{d^2k_{\perp}dk_3}\;.
\end{equation}

\section{Equivalence of different approaches in 1+1D}
\label{sec:III}
\subsection{Our approach}
From \eqref{III5}, which applies in 1+3D, we can deduce the in-mode in 1+1D straightforwardly by
\begin{equation}
\begin{aligned}
&u^{(1)}(t,x)\approx u_{h}(t,x)-\alpha\int_{\mathbb{R}} dt_m'dx'\;\gamma^{-1}(t_m')
\\
&\times \delta(x'-z_m(t_m'))u_h(t_m',x')G_{R}(t,x;t_m',x')\;,
\label{II1}
\end{aligned}
\end{equation}
where
\begin{equation}
\begin{aligned}
&G_{R}(t,x;t_m',x')=\frac{1}{2}\Theta(t-t_m'-|x-x'|)
\\
&=\frac{1}{2}\int_{-\infty}^{t}dt^{''}\delta(t^{''}-t_m'-|x-x'|)
\;,
\end{aligned}
\end{equation}
is the 1+1D retarded Green function, and we have changed the notations for the mirror's trajectory by $q(x_0')\rightarrow z_m(t_m')$, the observation points by $x_0\rightarrow t, x_3\rightarrow x$, and the dummy variables by $x_0'\rightarrow t_m',x_3'\rightarrow x'$ for a clear correspondence with the typical 1+1D literature.

For $u_h(t,x)=e^{-i\omega t-i\omega x}$ and on the mirror's right, i.e., $x-z_m(t_m')>0$, the inhomogeneous part of \eqref{II1} can be evaluated as
\begin{equation}
\begin{aligned}
&-\frac{\alpha}{2}\int_{-\infty}^{t}dt^{''}\int_{\mathbb{R}} dt_m'\gamma^{-1}(t_m')e^{-i\omega t_m'-i\omega z_m(t_m')}
\\
&\quad\times\delta(t^{''}-t_m'-x+z_m(t_m'))
\\
&=-\frac{\alpha}{2}\int_{-\infty}^{t}dt^{''}\int \frac{dR(t_m')}{1-\dot{z}_m(t_m')}\;\gamma^{-1}(t_m')
\\
&\quad\times e^{-i\omega t_m'-i\omega z_m(t_m')}\delta(t^{''}-x-R(t_m'))
\\
&=-\frac{\alpha}{2}\int_{-\infty}^{t_m(u)}dt_m'\gamma^{-1}(t_m')e^{-i\omega t_m'-i\omega z_m(t_m')}\;,
\end{aligned}
\label{II3}
\end{equation}
where $R(t_m')=t_m'-z_m(t_m')$ in the first equality and $R(t_m)=t-x$ in the last equality. Notice that $R(t_m)=t-x$ recovers the standard condition for an out-going photon in the null coordinates $u=t-x$, and hence we denote $t_m$ by $t_m(u)$. On the mirror's left, i.e., $x-z_m(t_m')<0$, we have
\begin{equation}
\begin{aligned}
&-\frac{\alpha}{2}\int_{-\infty}^{t_m(v)}dt_m'\gamma^{-1}(t_m')e^{-i\omega t_m'-i\omega z_m(t_m')}
\\
&=-\frac{\alpha}{2}\int_{-\infty}^{t_m(u)}dt_m'\gamma^{-1}(t_m')\;
\\
&\quad\times e^{i\omega[t_m+z_m(t_m)-t_m'-z_m(t_m')]}e^{-i\omega t-i\omega x}\;,
\end{aligned}
\end{equation}
instead, where $t_m(v)$ is determined by $t_m+z_m(t_m)=t+x$, which again recovers the standard condition $v=t+x$ for an in-coming photon.

For the out-mode, we use the advanced Green function
\begin{equation}
\begin{aligned}
&G_{A}(t,x;t_m',x')=\frac{1}{2}\Theta(t_m'-t-|x-x'|)
\\
&=-\frac{1}{2}\int_{\infty}^{t}dt^{''}\delta(t_m'-t^{''}-|x-x'|)
\;,
\end{aligned}
\end{equation}
in \eqref{II1} instead. Following the same procedure as above, we find, for $u_h(t,x)=e^{-i\omega t+i\omega x}$ and on the mirror's right, i.e., $x-z_m(t_m')>0$, the inhomogeneous part is
\begin{equation}
\begin{aligned}
-\frac{\alpha}{2}\int_{t_m(v)}^{\infty}dt_m'\gamma^{-1}(t_m')e^{-i\omega t_m'+i\omega z_m(t_m')}\;.
\end{aligned}
\label{II6}
\end{equation}
The other situations, e.g., $u_h(t,x)=e^{-i\omega t-i\omega x}$ and on the mirror's left, may be straightforwardly found by using the same procedure and thus we shall not repeat it here.

\subsection{Nicolaevici's approach}
The in-mode given by Nicolaevici \cite{1d}\cite{1e} is, e.g.,
\begin{equation}
V^R=e^{-i\omega v}-R^R(u)e^{-i\omega p(u)},\; V^L=T^L(v)e^{-i\omega v}\;,
\end{equation}
where the superscripts $R/L$ refer to the mirror's right/left, $u,v$ are the 1+1D null coordinates, and the ray-tracing function is
\begin{equation}
p(u)=2z_m(u)+u\;,
\end{equation}
and the reflection and transmission coefficients are
\footnote{Unitarity guarantees that $|R|^2+|T|^2=1$. However, some caution is required when one computes quantities such as $|R|^2+|T|^2=1$ and $R+T=1$ in the expansion of $\alpha$. For example, to verify $|R|^2+|T|^2=1$ to the accuracy of second order in $\alpha$, it is sufficient to retain up to the first order in $R$, whereas one must keep track to the second order in $T$. This is because the interference between the zeroth order and the second order of $T$ also contributes to $|T|^2$. In contrast, to check $R+T=1$, both $R$ and $T$ can be expanded to the same order since there is no interference involved.}
\begin{equation}
\begin{aligned}
R^R(u)&=\frac{\alpha}{2}\int_{-\infty}^{\tau}d\tau'\;e^{-\frac{\alpha}{2}(\tau-\tau')+i\omega[v(\tau)-v(\tau')]}\;,
\\
T^L(v)&=1-R^R(u)\;,
\end{aligned}
\end{equation}
where $\tau$ is the mirror's proper time and $v(\tau)=t_m(u)+z_m(t_m)=p(u)$. In the first-order approximation, which corresponds to the semi-transparent limit \footnote{During acceleration, the reflection coefficient may be manipulated as, e.g., $R^R(v)\sim\frac{\lambda}{2}\int_{\tau}^{\tau_A}d\tau'e^{\lambda(\tau-\tau')/2}\left(\cdots\right)=\frac{\lambda(\tau_A-\tau)}{2}\int_{0}^{1}d\sigma e^{-\lambda(\tau_A-\tau)\sigma/2}\left(\cdots\right)$, where we have made a change of variable by $\tau'-\tau=(\tau_A-\tau)\sigma$. It now becomes clear that the first-order approximation corresponds to the semi-transparent limit, i.e., $|R^R|\ll 1$. For more discussion on this, please see \cite{1c}\cite{1d}.}, the reflection coefficient becomes
\begin{equation}
R^R(u)\approx \frac{\alpha}{2}\int_{-\infty}^{\tau}d\tau'\;e^{i\omega[v(\tau)-v(\tau')]}\;.
\end{equation}
Therefore, we have
\begin{equation}
\begin{aligned}
&-R^R(u)e^{-i\omega p(u)}\approx -\frac{\alpha}{2}\int_{-\infty}^{\tau}d\tau'\;e^{-i\omega t_m'-i\omega z_m(t_m')}
\\
&=-\frac{\alpha}{2}\int_{-\infty}^{t_m(u)}dt_m'\gamma^{-1}(t_m')e^{-i\omega t_m'-i\omega z_m(t_m')}\;,
\end{aligned}
\end{equation}
which recovers our result \eqref{II3}.

The out mode is given by \cite{1d}\cite{1e}
\begin{equation}
U^R=e^{-i\omega u}-R^R(v)e^{-i\omega f(v)},\; U^L=T^L(u)e^{-i\omega u}\;,
\end{equation}
where the ray-tracing function is
\begin{equation}
f(v)=-2z_m(v)+v\;,
\end{equation}
and the reflection and transmission coefficients are
\begin{equation}
\begin{aligned}
R^R(v)&=\frac{\alpha}{2}\int_{\tau}^{\infty}d\tau'\;e^{\frac{\alpha}{2}(\tau-\tau')+i\omega [u(\tau)-u(\tau')]}\;,
\\
T^L(u)&=1-R^R(v)\;,
\end{aligned}
\end{equation}
where $u(\tau)=t_m(v)-z_m(t_m)=f(v)$. In the first-order limit, we obtain
\begin{equation}
\begin{aligned}
&-R^R(v)e^{-i\omega f(v)}
\\
&\approx-\frac{\alpha}{2}\int_{t_m(v)}^{\infty}dt_m'\gamma^{-1}(t_m')e^{-i\omega t_m'+i\omega'z_m(t_m')}\;,
\end{aligned}
\end{equation}
which is identical to our \eqref{II6}.

The beta-coefficients on the mirror's right using Nicolaevici's modes are
\begin{equation}
\begin{aligned}
&\beta_{\omega \omega'}^{ref}=-\left<U^{out*}(\omega>0),V^{in}(\omega'>0)\right>
\\
&=-\left[ \frac{\omega}{2\pi\sqrt{\omega \omega'}}\int_{-\infty}^{\infty}du\;R^{R}(u)e^{-i\omega'p(u)}e^{-i\omega u} \right]^{*}
\\
&\stackrel{IBP}{=}\frac{\alpha}{4\pi i\sqrt{\omega\omega'}}\int_{-\infty}^{\infty}du\frac{\gamma^{-1}(t_m)}{1-\dot{z}_m(t_m)}e^{i(\omega+\omega')t_m-i(\omega-\omega')z_m(t_m)}
\\
&=\frac{\alpha}{4\pi i\sqrt{\omega\omega'}}\int_{-\infty}^{\infty}dt_m\gamma^{-1}(t_m)e^{i(\omega+\omega')t_m-i(\omega-\omega')z_m(t_m)}\;, 
\end{aligned}
\end{equation}
where the superscript "\textit{ref}" refers to beta-coefficient due to reflected modes and "\textit{IBP}" refers to integration by parts, and note that $du=(1-\dot{z}_m(t_m))dt_m$. For the beta-coefficient due to the transmitted modes, the other set of in-mode is required, see \cite{1d}\cite{1e}. Since the discussion is similar, we simply list the result we obtained:
\begin{equation}
\begin{aligned}
&\beta_{\omega \omega'}^{tran}
\\
&=\frac{\alpha}{4\pi i}\frac{1}{\sqrt{\omega\omega'}}\int_{-\infty}^{\infty}dt_m\gamma^{-1}(t_m)e^{i(\omega+\omega')t_m-i(\omega+\omega')z_m(t_m)}\;.
\end{aligned}
\end{equation}
The above coefficients agree with the 1+1D limit of \eqref{III11} for $k_3'<0$ and $k_3'>0$, respectively.

We have now completed the demonstration of the equivalence between our approach and the literature's since we are able to obtain identical expressions for the mode functions and the beta-coefficients by further manipulating the standard expressions by an integration by part and a change of variable. Notice that the approach adopted and the expressions given in the standard literature are restricted to 1+1D since the analysis is based on the null coordinates. Nevertheless, our approach and expressions extend the discussion to higher dimensions.

\section{Gravitational Collapse}
\label{sec:IV}
We now apply our 1+3D formula, Eq.\eqref{III11}, to the trajectory that mimics the physics of gravitational collapse, which, up to now, has been investigated in 1+1D only. By comparing with the 1+1D results in the literature, we identify properties that are exclusive to higher, 1+3D dimensional spacetime. The trajectory of interest is
\begin{equation}
\begin{aligned}
&z_m(t_m)
\\
&=
\begin{cases}
0\;,\quad t_m\leq 0
\\
-t_m+\frac{1}{\kappa}-\frac{W[e^{1-2\kappa t_m}]}{\kappa}\;,\quad 0\leq t_m<\infty\;,
\end{cases}
\end{aligned}
\end{equation}
where $W(x)$ is the product logarithm and $\kappa$ is the surface gravity. The mirror is initially static and it begins to execute Carlitz-Willey(CW)-like acceleration after $t_m=0$. The following list the results for quantities in the acceleration phase that will appear in our later computation.
\begin{equation}
\begin{aligned}
&\frac{dz_m}{dt_m}=-\frac{1-W[e^{1-2\kappa t_m}]}{1+W[e^{1-2\kappa t_m}]}\;,
\\
&\gamma^{-1}(t_m)=\frac{2\sqrt{W[e^{1-2\kappa t_m}]}}{1+W[e^{1-2\kappa t_m}]}\;.
\end{aligned}
\end{equation}
On the mirror's right, the beta-coefficient due to the reflected mode may be evaluated by
\begin{align*}
&\beta_{\textbf{k}\textbf{k}'}^{ref}(k_3'>0)
\\
&\approx\frac{\alpha}{4\pi i}\frac{1}{\sqrt{|\textbf{k}||\textbf{k}'|}}\int_{-\infty}^{\infty}dt_m\gamma^{-1}(t_m)e^{i(|\textbf{k}|+|\textbf{k}'|)t_m-i(k_3-k_3')z_m(t_m)}
\\
&=-\frac{\alpha}{4\pi \sqrt{|\textbf{k}||\textbf{k}'|}}\left[\frac{1}{|\textbf{k}|+|\textbf{k}'|}\right]
\\
&
+\frac{\alpha}{4\pi i}\frac{1}{\sqrt{|\textbf{k}||\textbf{k}'|}}\int_{0}^{\infty}dt_m\gamma^{-1}(t_m)e^{i(|\textbf{k}|+|\textbf{k}'|)t_m-i(k_3-k_3')z_m(t_m)}.
\end{align*}
By making the change of variable:
\begin{equation}
\begin{aligned}
&d\zeta=\frac{2W[e^{1-2\kappa t_m}]}{1+W[e^{1-2\kappa t_m}]}dt_m
\\
&\zeta=\frac{1}{\kappa}-\frac{W[e^{1-2\kappa t_m}]}{\kappa},\quad t_m=\frac{\zeta}{2}-\frac{1}{2\kappa}\ln(1-\kappa\zeta)\;,
\end{aligned}
\end{equation}
we obtain
\begin{align*}
&\beta_{\textbf{k}\textbf{k}'}^{ref}(k_3'>0)\approx
-\frac{\alpha}{4\pi \sqrt{|\textbf{k}||\textbf{k}'|}}\left[\frac{1}{|\textbf{k}|+|\textbf{k}'|}\right]
\\
&+\frac{\alpha}{4\pi i}\frac{1}{\sqrt{|\textbf{k}||\textbf{k}'|}}\int_{0}^{\frac{1}{\kappa}}d\zeta\;(1-\kappa\zeta)^{-\frac{1}{2}-\frac{i}{2\kappa}(|\textbf{k}|+k_3+|\textbf{k}'|-k_3')}
\\
&\times e^{\frac{i}{2}(|\textbf{k}|-k_3+|\textbf{k}'|+k_3')\zeta}
\\
&=-\frac{\alpha}{4\pi \sqrt{|\textbf{k}||\textbf{k}'|}}\left[\frac{1}{|\textbf{k}|+|\textbf{k}'|}\right]+\frac{\alpha}{4\pi i\kappa }\frac{e^{\frac{i}{2\kappa}(|\textbf{k}|-k_3+|\textbf{k}'|+k_3')}}{\sqrt{|\textbf{k}||\textbf{k}'|}}
\\
&\times\int_{0}^{1}dz\;z^{-\frac{1}{2}-\frac{i}{2\kappa}(|\textbf{k}|+k_3+|\textbf{k}'|-k_3')}e^{-\frac{i}{2\kappa}(|\textbf{k}|-k_3+|\textbf{k}'|+k_3')z}\;,
\end{align*}
where $z=1-\kappa\zeta$. Next, performing a contour integration in the lower complex plane of $z$ and deforming the contour away from the pole $z=0$ (this small arc gives no contribution), we obtain
\begin{align*}
&\beta_{\textbf{k}\textbf{k}'}^{ref}(k_3'>0)\approx-\frac{\alpha}{4\pi \sqrt{|\textbf{k}||\textbf{k}'|}}\left[\frac{1}{|\textbf{k}|+|\textbf{k}'|}\right]
\\
&-\frac{\alpha}{4\pi \kappa \sqrt{|\textbf{k}||\textbf{k}'|}}\biggl[e^{\frac{i}{2\kappa}(|\textbf{k}|-k_3+|\textbf{k}'|+k_3')}e^{\frac{i\pi}{4}}e^{-\frac{\pi}{4\kappa}(|\textbf{k}|+k_3+|\textbf{k}'|-k_3')}
\\
&\times
\underbrace{\int_{0}^{\infty}ds\;s^{-\frac{1}{2}-\frac{i}{2\kappa}(|\textbf{k}|+k_3+|\textbf{k}'|-k_3')}e^{-\frac{(|\textbf{k}|-k_3+|\textbf{k}'|+k_3')}{2\kappa}s} }_{z=-is}
\\
&-\underbrace{\int_{0}^{\infty}ds\;(1-is)^{-\frac{1}{2}-\frac{i}{2\kappa}(|\textbf{k}|+k_3+|\textbf{k}'|-k_3')}e^{-\frac{(|\textbf{k}|-k_3+|\textbf{k}'|+k_3')}{2\kappa}s}  }_{z=1-is}\biggr]\;.
\end{align*}
The integrals can be evaluated in terms of Gamma and upper incomplete Gamma functions and the result is
\begin{equation}
\begin{aligned}
&\beta_{\textbf{k}\textbf{k}'}^{ref}(k_3'>0)\approx-\frac{\alpha}{4\pi \sqrt{\omega\omega'}}\left[\frac{1}{\omega+\omega'}\right]
\\
&-\frac{\alpha e^{\frac{i\omega_{-}^r}{2\kappa}}e^{\frac{i\pi}{4}}e^{-\frac{\pi\omega_{+}^r}{4\kappa}}}{4\pi \kappa \sqrt{\omega\omega'}}\left[\frac{2\kappa}{\omega_{-}^r}\right]^{\frac{1}{2}-\frac{i\omega_{+}^r}{2\kappa}}
\\
&\times
\biggl\{\Gamma\left[\frac{1}{2}-\frac{i\omega_{+}^r}{2\kappa} \right]-\Gamma\left[\frac{1}{2}-\frac{i\omega_{+}^r}{2\kappa},\frac{i\omega_{-}^r}{2\kappa}\right] \biggr\}\;,
\end{aligned}
\label{IV7}
\end{equation}
where we have defined $\omega_{+}^r=|\textbf{k}|+k_3+|\textbf{k}'|-k_3',\;\omega_{-}^r=|\textbf{k}|-k_3+|\textbf{k}'|+k_3'$, and $\omega=|\textbf{k}|=(\textbf{k}_{\perp}^2+k_3^2)^{1/2},\;\omega'=|\textbf{k}'|=(\textbf{k}_{\perp}^2+{k_3'}^2)^{1/2}$ for brevity. Following similar procedure, we obtain the beta-coefficient due to the transmitted modes as
\begin{equation}
\begin{aligned}
&\beta_{\textbf{k}\textbf{k}'}^{tran}(k_3'>0)\approx-\frac{\alpha}{4\pi \sqrt{\omega\omega'}}\left[\frac{1}{\omega+\omega'}\right]
\\
&-\frac{\alpha e^{\frac{i\omega_{-}^t}{2\kappa}}e^{\frac{i\pi}{4}}e^{-\frac{\pi\omega_{+}^t}{4\kappa}}}{4\pi \kappa \sqrt{\omega\omega'}}\left[\frac{2\kappa}{\omega_{-}^t}\right]^{\frac{1}{2}-\frac{i\omega_{+}^t}{2\kappa}}
\\
&\times
\biggl\{\Gamma\left[\frac{1}{2}-\frac{i\omega_{+}^t}{2\kappa} \right]-\Gamma\left[\frac{1}{2}-\frac{i\omega_{+}^t}{2\kappa},\frac{i\omega_{-}^t}{2\kappa}\right] \biggr\}\;,
\end{aligned}
\label{IV8}
\end{equation}
where $\omega_{+}^t=|\textbf{k}|+k_3+|\textbf{k}'|+k_3',\;\omega_{-}^t=|\textbf{k}|-k_3+|\textbf{k}'|-k_3'$.

The results \eqref{IV7} and \eqref{IV8} apply in 1+3D for $\textbf{k},\;\textbf{k}'\in\mathbb{D}$ satisfying the semi-transparent condition. However, we simply take this lower momentum cutoff for $k_3,k_3'$ as $k_c$ in this paper for simplicity.
\\
\\
$\blacksquare$ \textbf{Case 1: $\textbf{k}_{\perp}=0$ (1+1D limit)}

Perpendicular modes are effectively 1+1D. 
\\
Letting $\textbf{k}_{\perp}=0$ in \eqref{IV7} and \eqref{IV8} gives
\begin{equation}
\begin{aligned}
&\beta_{\omega\omega'}^{ref}\approx-\frac{\alpha}{4\pi\sqrt{\omega\omega'}}\left[\frac{1}{\omega+\omega'}\right]
\\&-\frac{\alpha e^{\frac{i\omega'}{\kappa}}e^{\frac{i\pi}{4}}}{4\pi \kappa \sqrt{\omega\omega'}}\left[\frac{\kappa}{\omega'}\right]^{\frac{1}{2}-\frac{i\omega}{\kappa}}e^{-\frac{\pi\omega}{2\kappa}}
\\
&\times
\biggl\{\Gamma\left[\frac{1}{2}-\frac{i\omega}{\kappa}\right]-\Gamma\left[\frac{1}{2}-\frac{i\omega}{\kappa},\frac{i\omega'}{\kappa}\right] \biggr\}\;,
\end{aligned}
\end{equation}
and
\begin{equation}
\begin{aligned}
&\beta_{\omega\omega'}^{tran}\approx-\frac{\alpha}{4\pi\sqrt{\omega\omega'}}\left[\frac{1}{\omega+\omega'}\right]
\\
&+\frac{\alpha}{4\pi i}\frac{1}{\sqrt{\omega\omega'}}\left[\frac{2}{\kappa-2i(\omega+\omega')} \right]\;.
\end{aligned}
\label{IV10}
\end{equation}

In the high frequency regime: $\omega'\gg\kappa$ for $\beta_{\omega\omega'}^{ref}$, using the asymptotic behavior for the upper incomplete Gamma function, i.e., $\Gamma(s,n)\approx n^{s-1}e^{-n}$ for $n\rightarrow\infty$, the third term exactly cancels out the first term in $\beta_{\omega\omega'}^{ref}$ by further assuming $\omega'\gg\omega$. The remaining contribution to $\beta_{\omega\omega'}^{ref}$ is thus the second term which gives
\begin{equation}
|\beta_{\omega \omega'}^{ref}|^2\approx\frac{\alpha^2}{8\pi\kappa\omega\omega'^2}\left[\frac{1}{e^{2\pi\omega/\kappa}+1}\right]\;,
\label{IV11}
\end{equation}
which reproduces the spectrum in \cite{1}\cite{1c}\cite{1d}. At this point, $\omega>k_c$ while $\omega'>k_c$, $\omega'\gg\kappa$, and $\omega'\gg\omega$. On the other hand, all the terms in \eqref{IV10} combine to give
\begin{equation}
\begin{aligned}
|\beta_{\omega \omega'}^{tran}|^2\approx\frac{\alpha^2\kappa^2}{16\pi^2\omega\omega'}\frac{1}{(\omega+\omega')^2[\kappa^2+4(\omega+\omega')^2]}\;.
\label{IV12}
\end{aligned}
\end{equation}
\\
$\blacksquare$ \textbf{Case 2: $\omega_{-}\ll\kappa$}

For $\omega_{-}^r\ll\kappa$, the first term in $\beta_{\textbf{k}\textbf{k}'}^{ref}$ dominates:
\begin{equation}
|\beta_{\textbf{k}\textbf{k}'}^{ref}(k_3'>0)|^2\approx \frac{\alpha^2}{16\pi^2\omega\omega'}\frac{1}{(\omega+\omega')^2}\;.
\end{equation}

For $0<\omega_{-}^t\ll\kappa$, the first term in $\beta_{\textbf{k}\textbf{k}'}^{tran}$ dominates:
\begin{equation}
|\beta_{\textbf{k}\textbf{k}'}^{tran}(k_3'>0)|^2\approx \frac{\alpha^2}{16\pi^2\omega\omega'}\frac{1}{(\omega+\omega')^2}\;.
\end{equation}
$\blacksquare$ \textbf{Case 3: $\omega'\gg\omega\wedge\omega'\gg\kappa$}

In this case, only the second term in $\beta_{\textbf{k}\textbf{k}'}^{ref}$ survives:
\begin{equation}
\begin{aligned}
&|\beta_{\textbf{k}\textbf{k}'}^{ref}(k_3'>0)|^2
\\
&\approx  \frac{\alpha^2}{8\pi\kappa\omega k_3'^2}\left[1-\frac{3(\omega\sin\theta)^2}{4k_3'^2}\right]\left[\frac{1}{e^{\omega/T_{\text{eff}}(\theta)}+1}\right]\;,
\end{aligned}
\label{IV15}
\end{equation}
where $T_{\text{eff}}(\theta)=\kappa/(1+\cos\theta)\pi$ is the effective temperature. At this point, the conditions required are: $k_3>k_c$, $k_3'>k_c$, $\omega'\sim k_3'\gg\omega$, and $\omega'\sim k_3'\gg\kappa$.

On the other hand, to expand the incomplete Gamma function for $\omega_{-}^t\gg\kappa$ in $\beta_{\textbf{k}\textbf{k}'}^{tran}$, the additional conditions: $\theta\neq 0$ and $\omega\gg\kappa$ are required. However, in such a case, the third term no longer cancels with the first term in $\beta_{\textbf{k}\textbf{k}'}^{tran}$ but only indicates the latter is negligible compared to the former. Nevertheless, since $\omega_{-}^t\gg\kappa$, the third term is negligible compared to the second term. Therefore, at the end of the day, the second term in $\beta_{\textbf{k}\textbf{k}'}^{tran}$ dominates and gives
\begin{equation}
\begin{aligned}
&|\beta_{\textbf{k}\textbf{k}'}^{tran}(k_3'>0)|^2\approx\frac{\alpha^2}{4\pi\kappa\omega^2k_3'}\left[\frac{e^{-2\pi k_3'/\kappa}}{1-\cos\theta}\right]\;,
\end{aligned}
\label{IV16}
\end{equation}
under the conditions: $k_3>k_c$, $k_3'>k_c$, $\omega'\sim k_3'\gg\omega$, $\omega'\sim k_3'\gg\kappa$, $\omega\gg\kappa$, and $\theta\neq 0$.

Using \eqref{III12}, \eqref{IV15}, and \eqref{IV16}, we are able to obtain analytic expressions for their respective particle spectra.

The reflected particle spectrum is
\begin{equation}
\begin{aligned}
&\frac{dN_{ref}(0\leq\theta\leq\pi/2)}{d\omega d\Omega}\approx\frac{A\omega^2}{4\pi^2}\int_{\Lambda_1}^{\infty} dk_3'\;|\beta_{\textbf{k}\textbf{k}'}^{ref}|^2
\\
&=\frac{A\alpha^2}{32\pi^3\kappa \Lambda_1}\left[1-\frac{(\omega\sin\theta)^2}{4\Lambda^2_1}\right]\left[\frac{\omega}{e^{\omega/T_{\text{eff}}(\theta)}+1}\right]\;,
\end{aligned}
\label{IV17}
\end{equation}
where $\Lambda_1=\kappa\vee\omega$ is the lower momentum cutoff.

The transmitted particle spectrum is
\begin{equation}
\begin{aligned}
&\frac{dN_{tran}(\theta=0)}{d\omega d\Omega}\approx\frac{A\omega^2}{4\pi^2}\int^{\infty}_{\Lambda_2} dk_3'\;|\beta_{\textbf{k}\textbf{k}'}^{tran}|^2 
\\
&\approx\frac{A\alpha^2}{1024\pi^4}\left[\frac{\kappa^2\omega}{\Lambda_2^4}\right]\;,
\\
&\frac{dN_{tran}(0<\theta\leq\pi/2)}{d\omega d\Omega}\approx\frac{A\omega^2}{4\pi^2}\int^{\infty}_{\Lambda_2} dk_3'\;|\beta_{\textbf{k}\textbf{k}'}^{tran}|^2 
\\
&\approx\frac{A\alpha^2}{16\pi^3\kappa}\left[\frac{\Gamma\left(0,2\pi\Lambda_2/\kappa\right)}{1-\cos\theta}\right]\;,
\end{aligned}
\label{IV18}
\end{equation}
where $\Lambda_2=\omega>k_c$ is the lower momentum cutoff.
\begin{figure}[!htb]
\minipage{0.35\textwidth}
  \includegraphics[width=\linewidth]{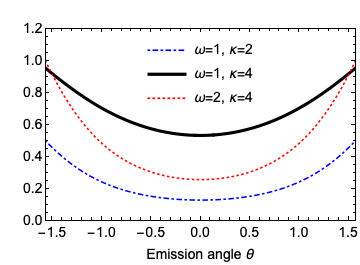}
  \caption{Angular spectrum for Eq.\eqref{IV17} when $\kappa>\omega$. The spectrum is normalized by the value at $\omega=2,\;\kappa=4$.}\label{fig:1a}
\endminipage\hfill
\minipage{0.35\textwidth}
  \includegraphics[width=\linewidth]{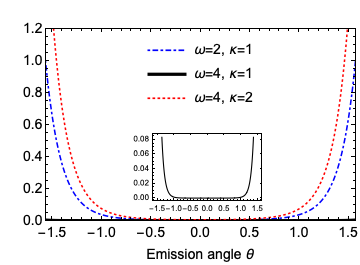}
  \caption{Angular spectrum for Eq.\eqref{IV17} when $\kappa<\omega$. The spectrum is normalized by the value at $\omega=4,\;\kappa=2$. The subgraph is a zoom-in for the case: $\omega=4,\kappa=1$ normalized by the value at $\theta=\pi/2$. }\label{fig:1b}
\endminipage\hfill
\minipage{0.35\textwidth}%
  \includegraphics[width=\linewidth]{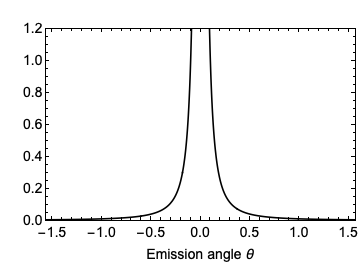}
  \caption{Angular spectrum for Eq.\eqref{IV18}. The spectrum is normalized by the value at $\theta=0.1$.}\label{fig:1c}
\endminipage
\end{figure}

In 1+3D spacetime, the number of particles emitted in the off-perpendicular directions due to the reflected modes is larger than those emitted perpendicularly to the mirror's surface, as illustrated in Fig.~\ref{fig:1a} and Fig.~\ref{fig:1b}. In addition, the motion of the mirror being relativistic is also crucial for this phenomenon to occur. This should be expected since, classically, the reflection of photons off a relativistic, receding mirror in 3-dimensional space tends to spread in large angles when the striking process is off-perpendicular \cite{8a}\cite{8b}. Therefore, in our current situation, there are more in modes reflected off-perpendicularly and thus the excitation of these modes leads to more off-perpendicular particles being created compared to their perpendicular counterpart. As for the particles created by the transmitted modes, they are mainly focused within a small emission angle, as illustrated in Fig.~\ref{fig:1c}. Thus, in 1+3D, the created perpendicular particles are the product of both the reflected, perpendicular in modes and the transmitted modes although the contribution from the latter may be negligible in comparison. However, for off-perpendicular directions, the created particles may serve as a characteristic product of the reflected in modes.

Furthermore, the effective temperature $T_{\text{eff}}(\theta)$ is emission angle ($\theta$)-dependent. In the perpendicular direction, i.e., effectively 1+1D, the effective temperature recovers the familiar $T_{\text{eff}}(\theta=0)=\kappa/2\pi$ in the 1+1D literature. However, as the emission angle gets larger, the effective temperature monotonically increases and eventually reaches twice the value of $T_{\text{eff}}(\theta=0)=\kappa/2\pi$ at $\theta=\pi/2$, i.e., $T_{\text{eff}}(\theta=\pi/2)=\kappa/\pi$. In fact, this tendency may be understood as a manifestation of the fact that off-perpendicular particles are more probable to be created as mentioned in the last paragraph.

\section{Conclusion}
\label{sec:V}
In this paper, we have considered the production of scalar particles by a relativistic, semi-transparent mirror based on the Barton-Calogeracos action and derived the corresponding particle spectrum in 1+3D and identify the relation between the spectrum and the particle production probability. Comparison of our treatment to the approach adopted in the literature in 1+1D is also demonstrated.

We apply our derived formula to the gravitational collapse trajectory in 1+3D. The spectra in various frequency/momentum regimes are derived analytically. In particular, in the regime $\omega'\gg\omega$ and $\omega'\gg\kappa$, we find the particle spectrum created by the reflected in modes has an effective temperature depending on the emission angle monotonically. In addition, there are more particles created with non-vanishing transverse momentum compared to the perpendicular ones due to the relativistic property of the mirror and the spacetime dimension being 1+3D.

In this paper, the mirror considered is an infinite-size, homogeneous, plane mirror in 1+3D Minkowski spacetime. However, the formalism adopted in principle allows the consideration of a mirror with finite size by, e.g., inserting a density function describing the mirror's transverse geometry. In addition, the geometric factor of the mirror is incorporated into the particle spectra, i.e., \eqref{IV17} and \eqref{IV18}, via the area $A$, which has a dimension of length${}^2$. If we group a factor of $\omega^2$ to the area $A$, they combine to give $A/\lambda^2\rightarrow\infty$, where $\lambda$ is the wavelength of the particle (another $\omega^2$ should be divided by $\alpha^2$ simultaneously giving the semi-transparent condition $\alpha/\omega\ll 1$). This observation indicates that the quantities we discussed are valid in the realm of geometric optics. When the finite-size effect is considered, the characteristic length $\sqrt{A}$ may be comparable to the wavelength $\lambda$. In such a case, diffraction may occur and the particle spectrum may include other corrections in terms of the characteristic length. The issue of finite-size effect will be further investigated in our upcoming work.

\begin{acknowledgments}
The authors appreciate helpful discussions with Yung-Kun Liu of National Taiwan University. This work is supported by ROC Ministry of Science and Technology (MOST), National Center for Theoretical Sciences (NCTS), and Leung Center for Cosmology and Particle Astrophysics (LeCosPA) of National Taiwan University. P.C. is in addition supported by U.S. Department of Energy under Contract No. DE-AC03-76SF00515. 
\end{acknowledgments}

\appendix
\section{S-matrix approach}
\label{A1}
In this appendix, we offer an alternative and fast way for computing the particle spectrum. This approach begins by recognizing the S-matrix of the BC action as
\begin{equation}
\mathbb{S}=\mathcal{T}e^{-\frac{i\alpha}{2}\int_{\mathbb{R}}d^4x\gamma^{-1}(x_0)\delta(x_3-q(x_0))\hat{\phi}_{I}^2(x)}\;,
\end{equation}
where $\mathcal{T}$ is the time-ordering operator and the subscript \textit{I} refers to the interaction picture.

By using the relation between the in-state/operator and the out-state/operator:
\begin{equation}
\begin{aligned}
\left|0,\text{out}\right>&=\mathbb{S}\left|0,\text{in}\right>\quad \& \quad \hat{a}_{\textbf{k}}^{\text{out}}=\mathbb{S}^{\dagger}\hat{a}_{\textbf{k}}^{\text{in}}\mathbb{S}\;,
\end{aligned}
\end{equation}
and the identity (for later convenience):
\begin{equation}
\begin{aligned}
&\left<0,\text{in}\right|\hat{a}_{\textbf{k}'}^{\text{in}}\hat{a}_{\textbf{p}}^{\text{in}}\hat{a}^{\text{in}\dagger}_{\textbf{k}}\hat{a}^{\text{in}}_{\textbf{k}}\hat{a}^{\text{in}\dagger}_{\textbf{p}'}\hat{a}^{\text{in}\dagger}_{\textbf{q}}\left|0,\text{in}\right>
\\
&=\delta(\textbf{k}'-\textbf{q})\delta(\textbf{p}-\textbf{k})\delta(\textbf{k}-\textbf{p}')
\\
&\quad+\delta(\textbf{k}'-\textbf{k})\delta(\textbf{k}-\textbf{p}')\delta(\textbf{p}-\textbf{q})
\\
&\quad+\delta(\textbf{k}'-\textbf{p}')\delta(\textbf{p}-\textbf{k})\delta(\textbf{q}-\textbf{k})
\\
&\quad+\delta(\textbf{k}'-\textbf{k})\delta(\textbf{q}-\textbf{k})\delta(\textbf{p}-\textbf{p}')\;,
\end{aligned}
\end{equation}
we may then compute the particle spectrum by
\begin{equation}
\begin{aligned}
&\frac{dN}{d^2k_{\perp}dk_3}=\left<0,\text{in}\right|\hat{a}^{\text{out}\dagger}_{\textbf{k}}\hat{a}^{\text{out}}_{\textbf{k}}\left|0,\text{in}\right>
\\
&=\left<0,\text{in}\right|\mathbb{S}^{\dagger}\hat{a}^{\text{in}\dagger}_{\textbf{k}}\hat{a}^{\text{in}}_{\textbf{k}}\mathbb{S}\left|0,\text{in}\right>
\\
&\approx\frac{A\alpha^2}{64\pi^4|\textbf{k}|}\int_{\mathbb{D}} dk_3'\frac{1}{\sqrt{k_{\perp}^2+{k_3'}^2}}
\\
&\times
\left|\int_{\mathbb{R}} dx_0'\gamma^{-1}(x_0')e^{i(|\textbf{k}|+\sqrt{k_{\perp}^2+{k_3'}^2})x_0'-i(k_3+k_3')q(x_0')}\right|^2,
\end{aligned}
\end{equation}
which agrees exactly with our previous result, Eq.\eqref{III12}. The advantage of the S-matrix approach is that it enables one to obtain the particle spectrum directly in a simpler manner without the need of finding the mode functions, Eq.\eqref{III5} and Eq.\eqref{III6}, first and performing laborious calculations.

\end{document}